\documentclass[review]{elsarticle}
\usepackage{float}
\restylefloat{table}

\usepackage[utf8]{inputenc}

\usepackage{mathtools}

\usepackage{times} 
\usepackage{amsmath} 
\usepackage{amssymb}  
\usepackage{amsthm}
\usepackage{booktabs,ragged2e}
\usepackage{multirow}
\usepackage[flushleft]{threeparttable}
\usepackage{xcolor}
\usepackage[english]{babel}
\usepackage{lipsum}
\usepackage{hyperref}
\usepackage{graphics} 
\usepackage{epsfig} 
\usepackage[ruled,linesnumbered]{algorithm2e}

\newtheorem{theorem}{Theorem}

\newcounter{definitioncounter}

\renewcommand{\thedefinitioncounter}{\arabic{definitioncounter}}

\newenvironment{definition}[1][Definition]
{\refstepcounter{definitioncounter} %
	\begin{trivlist}
		\item[\hskip \labelsep {\bfseries #1 \thedefinitioncounter}]}
	{\end{trivlist}}

\newcounter{assumptioncounter}

\renewcommand{\theassumptioncounter}{\arabic{assumptioncounter}}

\newenvironment{assumption}[1][Assumption]
{\refstepcounter{assumptioncounter} %
	\begin{trivlist}
		\item[\hskip \labelsep {\bfseries #1 \theassumptioncounter}]}
	{\end{trivlist}}

\def\figref#1{Fig.~\ref{#1}}
\def\secref#1{Section~\ref{#1}}

\SetKwInput{KwRequire}{Require}  %

\journal{  }

\biboptions{authoryear}

\begin{document}
\begin{frontmatter}

%
%
%
%

\title{Kalman Filter-Based Distributed Gaussian Process for Unknown Scalar Field Estimation in Wireless Sensor Networks}
\author[label1]{Jaemin Seo}
\ead{qkek1019@unist.ac.kr}
\author[label1]{Geunsik Bae}
\ead{baegs94@unist.ac.kr}
\author[label1]{Hyondong Oh\corref{cor1}}
\ead{h.oh@unist.ac.kr}

\cortext[cor1]{Corresponding author.}
\address[label1]{Department of Mechanical Engineering, Ulsan National Institute of Science and Technology (UNIST), Ulsan 44919, Republic of Korea}

\begin{abstract}
In this letter, we propose an online scalar field estimation algorithm of unknown environments using a distributed Gaussian process (DGP) framework in wireless sensor networks (WSNs).
While the kernel-based Gaussian process (GP) has been widely employed for estimating unknown scalar fields, its centralized nature is not well-suited for handling a large amount of data from WSNs. To overcome the limitations of the kernel-based GP, recent advancements in GP research focus on approximating kernel functions as products of E-dimensional nonlinear basis functions, which can handle large WSNs more efficiently in a distributed manner. However, this approach requires a large number of basis functions for accurate approximation, leading to increased computational and communication complexities.
To address these complexity issues, the paper proposes a distributed GP framework by incorporating a Kalman filter scheme (termed as K-DGP), which scales linearly with the number of nonlinear basis functions. Moreover, we propose a new consensus protocol designed to handle the unique data transmission requirement residing in the proposed K-DGP framework. This protocol preserves the inherent elements in the form of a certain column in the nonlinear function matrix of the communicated message; it enables wireless sensors to cooperatively estimate the environment and reach the global consensus through distributed learning with faster convergence than the widely-used average consensus protocol.
Simulation results demonstrate rapid consensus convergence and outstanding estimation accuracy achieved by the proposed K-DGP algorithm. 
The scalability and efficiency of the proposed approach are further demonstrated by online dynamic environment estimation using WSNs.
\end{abstract}

\begin{keyword}
	Sensor fusion \sep Distributed systems \sep Gaussian Process \sep Consensus Algorithm
\end{keyword}
\end{frontmatter}

\section{Introduction}
\label{sec1}
\begin{figure}[!b]
	\centering
	\includegraphics[width=\linewidth]{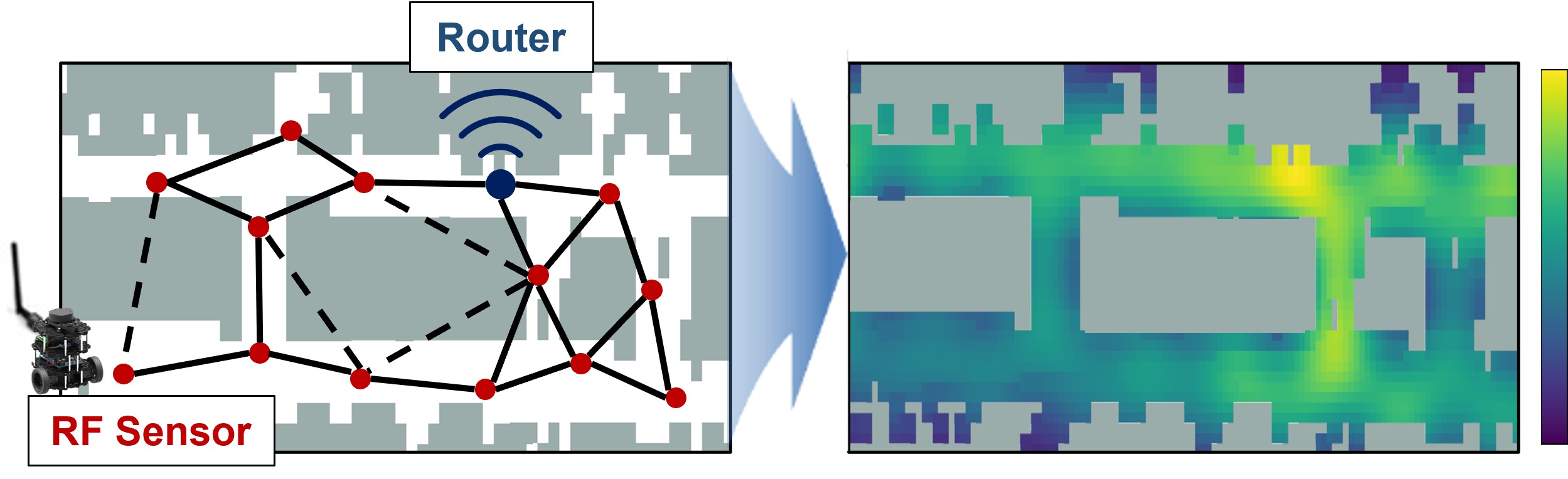}
	\caption{A sample scenario to predict communication channel using distributed wireless sensor networks. RF sensors are deployed as a distributed system to obtain communication signal strength measurements from a router. The proposed K-DGP algorithm is utilized to estimate a communication signal map (i.e., scalar field).}
	\label{distributed WSNs}
\end{figure}

Wireless sensor networks (WSNs) offer significant potential for effectively estimating unknown environments in a wide range of applications, such as surface reconstruction, environment monitoring, and surveillance \citep{kandris2020applications, domingo2016sensor}. An unknown environment refers to a scalar field where no prior information is available about the true underlying information distribution. In such scenarios, wireless sensors are deployed to gather data from the scalar field of interest and use the information to estimate a belief over the distribution (e.g., distance from obstacles \citep{SEO2023119459}, communication signals \citep{KIM20129}, and target likelihood \citep{KIM2109}) as illustrated in \figref{distributed WSNs}.

However, WSNs inherently face limitations in communication resources, making the use of a centralized framework for collecting data from sensors problematic. The centralized framework can lead to a bottleneck, resulting in delay, data loss and decreased reliability in scalar field estimation. To address this issue, distributed architectures could be considered \citep{julian2012distributed}. These methods can mitigate the bottleneck issue by enabling sensors to communicate and collaborate with one another, reducing reliance on a central node. Furthermore, integrating robust data fusion techniques and error-tolerant communication strategies can enhance the overall performance and reliability of WSNs in estimating unknown scalar fields \citep{9144385, fukami2021global}.

\subsection{Related Work}
Gaussian Process (GP) is widely used for estimating scalar fields due to its ability to handle complex relationships between sensor measurements and the inherent uncertainty of the data~\citep{wang2017online, guerrero2021sparse}. Specifically, the GP leverages kernel functions to estimate relationships within the data, which offers the advantage of enabling online estimation for time-varying unknown scalar fields when appropriate kernel functions are employed. However, these classic GP models with the centralized framework face challenges in effectively managing a large amount of data from WSNs due to their computational complexity. Moreover, all wireless sensors need to transmit data to a single central node for estimating the unknown scalar field, leading to high communication burdens among nodes. Although the use of communication relay to enhance the communication bandwidth is a known solution, it still consumes considerable time and may not be robust against communication failures~\citep{manfredi2013design}. As a result, significant efforts have been made to develop a distributed Gaussian Process (DGP) framework to address challenges in scalability of WSNs.
\cite{deisenroth2015distributed} proposed a hierarchical DGP approach using the kernel-based GP, which focused primarily on reducing computational complexity. However, this method still struggles with the increasing size of the measurements and it requires an additional computational step of re-estimation whenever the map resolution changes.

Recently, a new type of GP approach has emerged, which employs finite dimensional approximation techniques to reduce the computational burden of the kernel-based GP by approximating the kernel function with a sum of eigenfunction and eigenvalue product~\citep{8387507, solin2014hilbert}. This advancement has transformed the kernel-based classic GP framework into a basis-function-based framework, paving the way for efficient DGP algorithms. Notably in~\citep{9144385}, a basis-function-based GP combined with the average consensus algorithm, multi-agent DGP (MADGP), is applied in distributed multi-agent systems. In this approach, the transmitted data increases quadratically with $\mathcal{O}(E^2)$, where $E$ represents the number of functions used to approximate the kernel. This relationship changes the computational complexity and communication load dependence of the GP from the number of sensor measurements to the number of functions $E$. However, accurately approximating the kernel requires a large number of eigenfunctions, which still imposes a significant computational and communication load. Furthermore, since finding an analytic form of basis functions for a time-varying kernel is challenging, it is difficult to apply this approximated DGP algorithm in estimating dynamic scalar fields.

As an alternative approach to reduce the computational burden of the GP dealing with all collected measurement data up to the current time step, a new approach called Kalman filter-based GP (K-GP) (or similarly recursive GP) method is introduced~\citep{solin2018modeling}. This method allows for the sequential estimation of unknown scalar fields within the Bayesian framework. Unlike the conventional GP, which relies on all historical data, the K-GP only requires data from the current time step, utilizing the Kalman filter's ability to process past information based on the Markov property. However, it has been only applied in single agent scenarios~\citep{8880505, viset2022extended}.
Besides, a simple extension of the K-GP to multiple agents could result in the high computational and communication complexities found in existing DGP methods.

\subsection{Contributions}
In this study, we make several key contributions to enhance the performance of the DGP in WSNs. The main contributions of our work are as follows:
\begin{enumerate}
	\item A new recursive DGP algorithm is proposed by extending the K-GP algorithm for distributed multi-agent systems. The extension is necessary since performance improvements are challenging to achieve with traditional approaches. This strategic refinement makes the algorithm effectively decrease its dependency on the number of eigenfunctions. Consequently, these improvements contribute to a reduction in both computational and communication complexities compared with existing DGP algorithms;
	
	\item In the proposed DGP framework, the data that each wireless sensor needs to exchange possesses a unique structure that includes intrinsic elements. These elements are key components assigned to specific positions (column) within the data matrix. To leverage this feature, we introduce a new consensus protocol, termed as \textit{dual-extrema consensus}, which enables the distributed sensors to quickly reach consensus by either maximizing or minimizing the intrinsic elements; and
	
	\item The proposed DGP approach demonstrates the capability to efficiently estimate both stationary and dynamic scalar fields in a distributed manner. It is worthwhile noting that the DGP approach for dynamic scenarios has been rarely reported to our best knowledge. This is achieved by incorporating the Kalman filter prediction concept while significantly enhancing its practicality for real-world problem-solving.
	
\end{enumerate}

\section{Problem Statement and Backgrounds} \label{sec2}
This section describes the basic concept for estimating a belief map of an unknown scalar field using the GP. At sensing time step $k$, each wireless sensor $r\in \mathcal{N}=\{1,\:\dots\:,R\ | R=\#$ of sensors$\}$ measures data $y_k^r \in \mathbb{R}$ of the unknown scalar field $f(\cdot)$ at a location $\textbf{x}_k^r \in \mathbb{R}^2$ with a noise $\epsilon \sim \mathit{N}(0, \sigma_n^2)$ represented as:
\vspace{-1mm}
\begin{equation}
	{y_k^r} = f(\textbf{x}_k^r) + \epsilon. \label{measurement}
\end{equation}
\vspace{-1mm}\noindent
The unknown scalar field $f(\cdot)$ is estimated with obtained data $D_{1:k}=\{(\textbf{X}_i, \textbf{y}_i) | i\in\{1,\: \dots\: ,k\}\},$ where $\textbf{X}_i\in\{ {\textbf{x}_i^1},\: \dots \:,{\textbf{x}_i^R}\}$ and
${\textbf{y}_ i\in\{y_i^1,\: \dots\: ,y_i^R\}} $. 

In order for all wireless sensors in distributed WSNs to estimate the same belief map, they need to exchange data by communicating with their neighboring sensors. The topology of WSNs is represented using an undirected graph $\mathcal{G}=\{V,\mathcal{E}\}$, where $V=\{v_r\:|\: r\in\mathcal{N}\}$ denotes the set of nodes $v_r$ and $\mathcal{E}=\{(v_i, v_j)\:|\: i,j \in \mathcal{N}, \: \lVert \textbf{x}_i - \textbf{x}_j \rVert \leq d_{comm} \}$ denotes the set of edges. The edge $(v_i, v_j)$ is assigned a value of $1$ if $\{\lVert \textbf{x}_i - \textbf{x}_j \rVert \leq d_{comm}\}$, and $0$ otherwise. The neighbor set of a wireless sensor $r$ is represented as $\mathcal{N}_r=\: \{j | \mathcal{E}_{\{r,j\}}=1, j\in\mathcal{N}\}$.

\subsection{Gaussian Process} \label{classic GP}

The GP estimates the unknown scalar field $f(\cdot)$ by modelling the relationship between queried inputs $\textbf{x}^{*}$ and outputs ${y}^{*}$ from the sensor data $D_{1:k}$ with a predefined kernel function~\citep{rasmussen2004gaussian}. The predicted distribution $p(y^*|D_{1:k}, \textbf{x}^*)\sim\mathit{N}(\mu_k(\textbf{x}^*), \Sigma_k (\textbf{x}^*))$ is represented by:
\vspace{-1.5mm}
\setlength{\arraycolsep}{0.0em}
\begin{eqnarray}
	\mu_k(\textbf{x}^*)&=&{K(\textbf{x}^*,\textbf{X})}{\left(K(\textbf{X},\textbf{X})+\sigma_n^2\mathbf{I}\right)}^{-1}\textbf{y}, \label{gp_mu}\\ 
	\Sigma_k(\textbf{x}^*)&=&K(\textbf{x}^*,\textbf{x}^*)
	-{K(\textbf{x}^*,\textbf{X})}{\left(K(\textbf{X},\textbf{X})+\sigma_n^2\mathbf{I}\right)}^{-1}K(\textbf{X},\textbf{x}^*),
	\label{gp_var}
\end{eqnarray}
\vspace{-1.5mm}\noindent
where $\textbf{X}=\textbf{X}_{1:k}$ and $\textbf{y}=\textbf{y}_{1:k}$ and $K(\cdot, \cdot)$ is a kernel function. The squared exponential kernel with its hyperparameters $\sigma_s$, and $l$ is the widely-used kernel in most applications, represented as:
\vspace{-1.5mm}
\begin{equation}
	K(\textbf{x},\textbf{x}')=\sigma_s^2 \exp \left( {\lVert \textbf{x} - \textbf{x}'\rVert}/{l^2} \right).
	\label{sekern}
\end{equation}
\vspace{-1.5mm}\noindent
As demonstrated in \eqref{gp_mu}$\sim$\eqref{gp_var}, the centralized kernel-based GP requires collection of all measurements $D_{1:k}$ to ensure that all wireless sensors have the same estimated results, which is called batch estimation. Consequently, the computational complexity of the kernel-based GP for scalar field estimation becomes $\mathcal{O}({(Rk)}^3)$, predominantly due to the inverse matrix computation, which grows dramatically as the number of measurements increases. Furthermore, the inherent characteristic of this conventional GP necessitates the use of a central node for the collection of data. This requirement notably intensifies the bottleneck risk, making the transformation to a distributed framework challenging.

\subsection{Multi-agent distributed Gaussian Process} \label{MADGP sec}

To address the issues of the kernel-based GP for the distributed system, nonlinear function-based GP methodologies have been proposed. Among various methodologies, the Kahunen-Loeve (KL) kernel expansion~\citep{Levy2008} is frequently employed to approximate kernel functions with its ability to decompose complex kernels into manageable parts.
This approximated kernel function is used for the multi-agent distributed GP (MADGP)~\citep{8387507, 9144385} as it is independent of the number of measurements; this algorithm is briefly explained in the following. 

First of all, the nonlinear function-based GP represents a kernel function as a sum of the products of eigenfunctions $\phi_e(\cdot)$ and their corresponding eigenvalues $\lambda_e$:
\vspace{-1.5mm}
\setlength{\arraycolsep}{0.0em}
\begin{eqnarray}
	K(\textbf{x},\textbf{x}')= \mbox{$\sum_{e=1}^{E}{\lambda_e \phi_e(\textbf{x}) \phi_e(\textbf{x}')}	+\sum_{e=E+1}^{\infty}{\lambda_e \phi_e(\textbf{x}) \phi_e(\textbf{x}')}$}.  \label{kernel_sum}
\end{eqnarray}
\vspace{-1.5mm}\noindent
Although finding an analytic form of the eigenfunctions and eigenvalues can be challenging, the KL kernel expansion of the squared exponential has been well-studied~\citep{rasmussen2004gaussian}. As indicated in~\citep{8387507}, if $\lambda_e$ is approximately equal to $0$ for $e>E$ in the second term on the right-hand side of \eqref{kernel_sum}, the sum of the products of the first E-dimensional functions can provide a sufficient approximation result for the kernel. Therefore, we can represent the kernel function as the sum of the products of E-dimensional eigenfunctions and eigenvalues as:
\vspace{-1.5mm}
\setlength{\arraycolsep}{0.0em}
\begin{eqnarray}
	K(\textbf{x},\textbf{x}')\simeq \widehat{K}(\textbf{x},\textbf{x}')= \mbox{$\sum_{e=1}^{E}{\lambda_e \phi_e(\textbf{x}) \phi_e(\textbf{x}')}$}. \label{kernel_apx}
\end{eqnarray}
\vspace{-1.5mm}\noindent
By utilizing the approximated kernel function \eqref{kernel_apx}
and applying the matrix inversion lemma, the predicted distribution \mbox{$p(y^*|D_{1:k}, \textbf{x}^*)\sim\mathit{N}\left(\widehat{\mu}_k(\textbf{x}^*), \widehat{\Sigma}_k (\textbf{x}^*)\right)$} can be expressed as:
\vspace{-1.5mm}
\setlength{\arraycolsep}{0.0em}
\begin{eqnarray}
	\widehat{\mu}_k(\textbf{x}^*)&=& \mathsf{\Phi}^\mathrm{T}_*{\left( {\frac{{\boldsymbol{\Phi}}\boldsymbol{\Phi}}{Rk}}^\mathrm{T}+\frac{\sigma_n^2}{Rk}\Lambda^{-1}\right)}^{-1}\frac{\boldsymbol{\Phi}\textbf{y}}{Rk}, \label{MADGP mean}\\ 
	\widehat{\Sigma}_k(\textbf{x}^*)&=&\mathsf{\Phi}^\mathrm{T}_*{\left({\frac{{\boldsymbol{\Phi}}\boldsymbol{\Phi}}{Rk}}^\mathrm{T}+\frac{\sigma_n^2}{Rk}\Lambda^{-1}\right)}^{-1}\mathsf{\Phi}_*, \label{MADGP var}
\end{eqnarray}
\vspace{-1.5mm}\noindent
where $\mathsf{\Phi}_*=\mathsf{\Phi}(\textbf{x}^*)$, $\Lambda=diag\left(\lambda_1,\;\dots\;,\lambda_E\right)$, and
\setlength{\arraycolsep}{0.0em}
\vspace{-1mm}
\begin{eqnarray}
	{\mathsf{\Phi}(\textbf{x})}&=&[\phi_1(\textbf{x}),\;\dots\;,\phi_E(\textbf{x})]^\mathrm{T},\label{phi}\\
	{\boldsymbol{\Phi}\boldsymbol{\Phi}^\mathrm{T}}&=&\mbox{$\sum_{r=1}^{R}{\sum_{i=1}^{k}{\mathsf{\Phi}(\textbf{x}_i^r){\mathsf{\Phi}^\mathrm{T}(\textbf{x}_i^r)}}}$},\label{alpha}\\
	{\boldsymbol{\Phi}\textbf{y}}&=&\mbox{$\sum_{r=1}^{R}{\sum_{i=1}^{k}{\mathsf{\Phi}(\textbf{x}_i^r)y_i^r}}$}.\label{beta}
\end{eqnarray}
\vspace{-1.5mm}
\setlength{\arraycolsep}{5pt}\noindent
The predicted distribution \eqref{MADGP mean}$\sim$\eqref{MADGP var} can be reformulated as the following distributed form for each sensor: 
\setlength{\arraycolsep}{0.0em}
\begin{eqnarray}
	\widehat{\mu}_k^r(\textbf{x}^*)&=&\mathsf{\Phi}^\mathrm{T}_*{\left({\alpha}_k^r (0)+\frac{\sigma_n^2}{Rk}\Lambda^{-1}\right)}^{-1}{\beta}_k^r(0), \label{MADGP_mu}\\ 
	\widehat{\Sigma}_k^r(\textbf{x}^*)&=&\mathsf{\Phi}^\mathrm{T}_*{\left({\alpha}_k^r(0)+\frac{\sigma_n^2}{Rk}\Lambda^{-1}\right)}^{-1}\mathsf{\Phi}_*, 	\label{MADGP_var}
\end{eqnarray}
\setlength{\arraycolsep}{5pt}\noindent
where
\setlength{\arraycolsep}{0.0em}
\begin{eqnarray}
	{\alpha}_k^r(0) &=&\frac{k-1}{k} {\alpha}_{k-1}^r(0)+\frac{1}{k} \mathsf{\Phi}(\textbf{x}_k^r)\mathsf{\Phi}^\mathrm{T}(\textbf{x}_k^r), \label{alpha_r}\\
	{\beta}_k^r(0)&=&\frac{k-1}{k} {\beta}_{k-1}^r(0)+\frac{1}{k} \mathsf{\Phi}(\textbf{x}_k^r)y_k^r. \label{beta_r}
\end{eqnarray}
\setlength{\arraycolsep}{5pt}\noindent
In a distributed network, each sensor communicates both \eqref{alpha_r} and \eqref{beta_r} with its neighbor sensors only. The widely-used average consensus algorithm enables every node in the network to reach a consensus on the average of their initial values asymptotically~\citep{1239709}:
\setlength{\arraycolsep}{0.0em}
\begin{eqnarray}
	\lim_{t\rightarrow \infty} {\alpha_k^r}(t) &=& \mbox{$\frac{1}{R} \sum_{r=1}^{R}{{\alpha_k^r(0)}}, \:\:\:\: \forall r \in \mathcal{N}$},\label{avg1} \\
	\alpha_k^r(t+1) &=& \alpha_k^r(t) - \gamma \mbox{$\sum_{j\in \mathcal{N}}[\alpha_k^r(t) - \alpha_k^j(t)]$}, \label{avg2}
\end{eqnarray}
\setlength{\arraycolsep}{5pt}\noindent
where $\gamma$ is a consensus coefficient.
By applying the widely-used average consensus protocol in \eqref{avg2} to \eqref{alpha_r}$\sim$\eqref{beta_r}, \eqref{MADGP_mu}$\sim$\eqref{MADGP_var} can be converged to \eqref{MADGP mean}$\sim$\eqref{MADGP var} after sufficient consensus iterations $t$.

\begin{figure}[t]
	\centering
	\includegraphics[width=\linewidth, height=0.23\columnwidth]{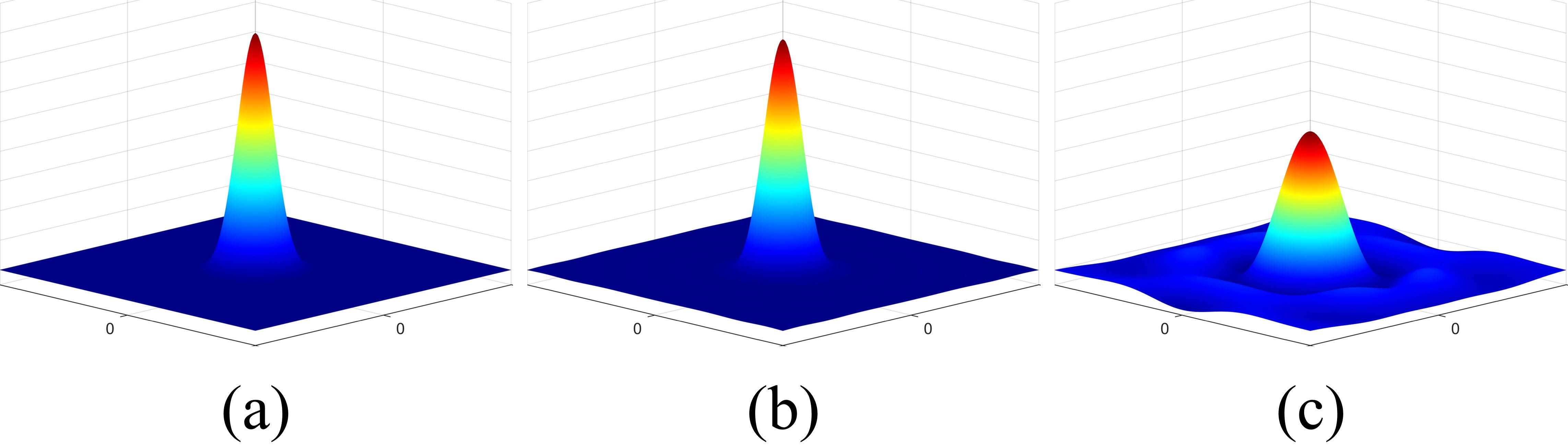}
	\caption{Approximation results of the kernel function based on the number of nonlinear functions $E$. (a) The squared exponential kernel with $\sigma_s=4$ and $l=0.07$, (b) E-dimensional kernel approximation $(E=400)$ and (c) E-dimensional kernel approximation $(E=80)$.}
	\label{kernel_approx}
\end{figure}
Unlike the kernel-based GP, which has a complexity that depends on the number of collected measurements $(Rk)$, the complexity of the MADGP relies only on the number of eigenfunctions $(E)$, making it a more scalable solution for large-scale networks. However, accurate kernel function approximation requires high-dimensional nonlinear functions (i.e., a large value of $E$). The results of approximating the kernel function in \eqref{sekern} with different values $E$ are illustrated in \figref{kernel_approx}. When a relatively small value $E$ is employed, it results in an imprecise approximation of the kernel function as shown in \figref{kernel_approx}(c). On the other hand, employing a larger value $E$ produces more accurate approximation, but it increases both the computational complexity $\mathcal{O}(E^3)$ associated with matrix inversion and the message size $\mathcal{O}(E^2)$ required for communication; this makes it difficult to use the MADGP framework for accurate scalar field estimation in real-time.

\section{Kalman Filter-Based Distributed~Gaussian~Process} \label{sec3}
As previously explained, the batch estimation of a conventional GP has computational challenges due to the requirement of using all collected measurements up to the current time step. Besides, the nonlinear function-based GP methods might also be computationally expensive when using a large value of $E$. In order to address these issues, a recursive Bayesian estimation approach for a single agent has been proposed~\citep{solin2018modeling}, which estimates unknown scalar fields sequentially with the nonlinear function-based GP. For the distributed K-GP, we found that the information filter, a commonly-used algorithm for Kalman filter decentralization, does not alleviate computational or communication burdens. To address this, we introduced a measurement matrix with a unique structure, designed so that only specific columns hold valid values for each sensor; this significantly reduces both computational and communication needs. Furthermore, we proposed a novel protocol that ensures all sensors achieve consistent estimation results, outperforming the average consensus protocol in both efficiency and accuracy.
The proposed algorithm not only uses less computation and smaller message size compared with the MADGP, but also preserves the scalability of the nonlinear function-based GP.

\subsection{GP with recursive Bayesian estimation} \label{sec3a}
We first briefly introduce the Kalman filter-based GP (K-GP) for a single agent~\citep{solin2018modeling}.
From the Bayesian nonlinear regression model with the eigenfunctions $f(\textbf{x})=\textbf{w}^\mathrm{T} \mathsf{\Phi}(\textbf{x})$, the original inference of the batch estimation in \secref{classic GP} over all possible parameters~$\textbf{w}$ is represented as:
\setlength{\arraycolsep}{0.0em}
\begin{eqnarray}
	p(y^*|D_{1:k}, \textbf{x}^*)&=&\int{p(y^*|\textbf{x}^*, \textbf{w})p(\textbf{w}|D_{1:k})\, d\textbf{w}}. \label{batch GP}	
\end{eqnarray}
\setlength{\arraycolsep}{5pt}\noindent
To obtain a recursive solution to the batch estimation \eqref{batch GP} for a single sensor $r$, 
$p(\textbf{w}|D_{1:k}^r)$ is initially estimated using the Bayes' rule according to~\cite{sarkka2013bayesian}, represented as: 
\setlength{\arraycolsep}{0.0em}
\begin{equation}
	p(\textbf{w}|D_{1:k}^r)\propto p(y_k|\textbf{w}, \textbf{x}_k)p(\textbf{w}|D_{1:k-1}^r)
	\propto N(\textbf{w}|\textbf{m}_k, \textbf{P}_k),  \label{single w}
\end{equation}
\setlength{\arraycolsep}{5pt}\noindent
where $\textbf{m}_k \in\mathbb{R}^{E\times1}$ and $\textbf{P}_k \in \mathbb{R}^{E\times E}$,
\setlength{\arraycolsep}{0.0em}
\begin{eqnarray}
	\textbf{m}_k&=&\left[\textbf{P}^{-1}_{k-1}+\frac{1}{\sigma_n^2}\mathsf{\Phi}(\textbf{x}_k^r)\mathsf{\Phi}^{\mathrm{T}}(\textbf{x}_k^r)\right]^{-1}\left[\frac{1}{\sigma_n^2}\mathsf{\Phi}(\textbf{x}_k^r)y_k^r + \textbf{P}^{-1}_{k-1}\textbf{m}_{k-1}\right], \label{KF_pre_mean}\\
	\textbf{P}_k&=&\left[\textbf{P}^{-1}_{k-1}+\frac{1}{\sigma_n^2}\mathsf{\Phi}(\textbf{x}_k^r)\mathsf{\Phi}^{\mathrm{T}}(\textbf{x}_k^r)\right]^{-1}. \label{KF_pre_var}
\end{eqnarray}
\setlength{\arraycolsep}{5pt}\noindent
By applying the matrix inversion lemma and introducing temporary variables $S_k$ and $\textbf{K}_k$, Equations \eqref{KF_pre_mean}$\sim$\eqref{KF_pre_var} can be represented as:
\setlength{\arraycolsep}{0.0em}
\begin{eqnarray}
	S_k &=& \mathsf{\Phi}^\mathrm{T}(\textbf{x}_k)\textbf{P}_{k-1}\mathsf{\Phi}(\textbf{x}_k)+\sigma_n^2, \label{KF_S} \\
	\textbf{K}_k &=& \textbf{P}_{k-1} \mathsf{\Phi}(\textbf{x}_k)S_k^{-1}, \label{KF_K}\\
	\mathbf{m}_k&=&\mathbf{m}_{k-1} + \mathbf{K}_k [y_k - \mathsf{\Phi}^\mathrm{T}(\textbf{x}_k)\mathbf{m}_{k-1}], \label{KF_m}\\
	\mathbf{P}_k &=&\mathbf{P}_{k-1}-\mathbf{K}_k S_k \mathbf{K}_k^\mathrm{T}. \label{KF_P}
\end{eqnarray}
\setlength{\arraycolsep}{5pt}\noindent
As a type of the nonlinear function-based GP methods,~\cite{solin2018modeling} employed the Hilbert space methods for approximating the kernel function~\citep{solin2014hilbert} to build eigenfunctions $\phi_e$ and eigenvalues $\lambda_e$, initializing $\mathbf{m}_0 = \mathbf{0}$ and $\mathbf{P}_0 =\overline{\Lambda}$ with
\begin{eqnarray}	
	\overline{\Lambda}=diag\left(S(\lambda_1),\;\dots\;,S(\lambda_E)\right), \label{lambdaS}
\end{eqnarray}
where $S(\cdot)$ denotes the spectral density of the kernel function required for the Hilbert space methods. It is worthwhile noting that, this mathematical formulation can be considered as a special case of the Kalman filter without the prediction step for a stationary scalar field estimation.

In order to transform above single agent recursive estimation into multi-agent estimation, that is, the Kalman filter-based DGP (K-DGP), the first concept we can consider is the information filter. The information filter~\citep{olfati2009kalman}, a type of the distributed Kalman filter widely used in WSNs, requires calculating an inverse matrix of \eqref{KF_P} to get the information matrix which should be shared with neighbor sensors. Consequently, this leads to the same computational complexity and message size as seen in the MADGP. To further reduce the computational complexity and the message size for communication, we introduce a modified measurement matrix $\mathbf{H} \in \mathbb{R}^{E\times R}$, replacing $\mathsf{\Phi}$ in \eqref{KF_S}$\sim$\eqref{KF_P}:
\setlength{\arraycolsep}{0.0em}
\begin{eqnarray}
	S_k &=& \mathbf{H}^\mathrm{T}\textbf{P}_{k-1}\mathbf{H}+\sigma_n^2 \mathbf{I}, \label{KF_S_mul} \\
	\textbf{K}_k &=& \textbf{P}_{k-1} \mathbf{H} S_k^{-1}, \label{KF_K_mul}\\
	\mathbf{m}_k&=&\mathbf{m}_{k-1} + \mathbf{K}_k [\textbf{y}_k - \mathbf{H}^\mathrm{T}\mathbf{m}_{k-1}], \label{KF_m_mul}\\
	\mathbf{P}_k &=&\mathbf{P}_{k-1}-\mathbf{K}_k S_k \mathbf{K}_k^\mathrm{T}, \label{KF_P_mul}
\end{eqnarray}
\setlength{\arraycolsep}{5pt}\noindent
where
\setlength{\arraycolsep}{0.0em}
\begin{eqnarray}
	\mathbf{H}&=&
	\begin{bmatrix}
		\phi_1(\textbf{x}_k^1) & \cdots & \phi_1(\textbf{x}_k^R) \\
		\vdots &\cdots & \vdots \\
		\phi_E(\textbf{x}_k^1) & \cdots & \phi_E(\textbf{x}_k^R) 
	\end{bmatrix}. \label{Multi_phi}
\end{eqnarray}
\setlength{\arraycolsep}{5pt}\noindent

Now, we replace $p(\textbf{w}|D_{1:k}^r)$ of \eqref{single w} by $p(\textbf{w}|D_{1:k})$ to determine the predicted distribution~$p(y^*|D_k, \textbf{x}^*) \sim \mathit{N}(\widehat{\mu}_k(\textbf{x}^*), \widehat{\Sigma}_k (\textbf{x}^*))$. By using the variance formula for linear transformation~\citep{Larry2003}, the mean and variance are represented as:
\vspace*{-2mm}
\setlength{\arraycolsep}{0.0em}
\begin{eqnarray}
	\widehat{\mu}_k(\textbf{x}^*)&=&\mathbb{E}[y^*|\textbf{w}]\approx \mathsf{\Phi}^\mathrm{T}_*\textbf{m}_k, \label{KF_mean} \\
	\widehat{\Sigma}_k (\textbf{x}^*)&=&\mathbb{V}[y^*|\textbf{w}]\approx \mathsf{\Phi}^\mathrm{T}_*\textbf{P}_k \mathsf{\Phi}_*. \label{KF_var}
\end{eqnarray}
\setlength{\arraycolsep}{5pt}
\vspace*{-6mm}

\noindent
For the K-DGP formulation, we decompose the modified measurement matrix $\mathbf{H}$ as:
\vspace*{-2mm}
\setlength{\arraycolsep}{0.0em}
\begin{eqnarray}
	\mathbf{H}&=&\mathbf{H}_1 + \cdots + \mathbf{H}_R, 
\end{eqnarray}
\setlength{\arraycolsep}{0pt}
\vspace*{-5mm}

\noindent 
where
\vspace*{-2mm}
\setlength{\arraycolsep}{0.0em}
\begin{eqnarray}
	\mathbf{H}_r&=&
	\begin{bmatrix}
		\Psi_r^1 & \cdots & \Psi_r^R
	\end{bmatrix}, \label{H}\\
	\Psi_r^n &=&
	\begin{cases}
		\mathsf{\Phi}(\textbf{x}_k^r), & \mbox{if}\; n=r, \\
		\textbf{0}, & otherwise,
	\end{cases}
	\forall n \in \mathcal{N}.
\end{eqnarray}
\setlength{\arraycolsep}{5pt}
\vspace*{-4mm}

\noindent
Here, $\mathbf{H}_r$ represents the message that needs to be communicated among neighbor sensors, characterized by a $\textbf{0}$ vector with the exception of its $r$-th column. This characterization is also applicable to $\textbf{y}_k$. In this framework, a key assumption is that each sensor is aware of the total number of sensors, denoted as $R$, and its pre-assigned individual sensor ID, $r$, which are essential for constructing matrix \eqref{H}. Although these assumptions might be restrictive for a distributed setting, it is noteworthy that the same conditions are also employed in the MADGP.

\begin{figure}[t]
	\centering
	\includegraphics[width=0.8\linewidth, height=0.60\columnwidth]{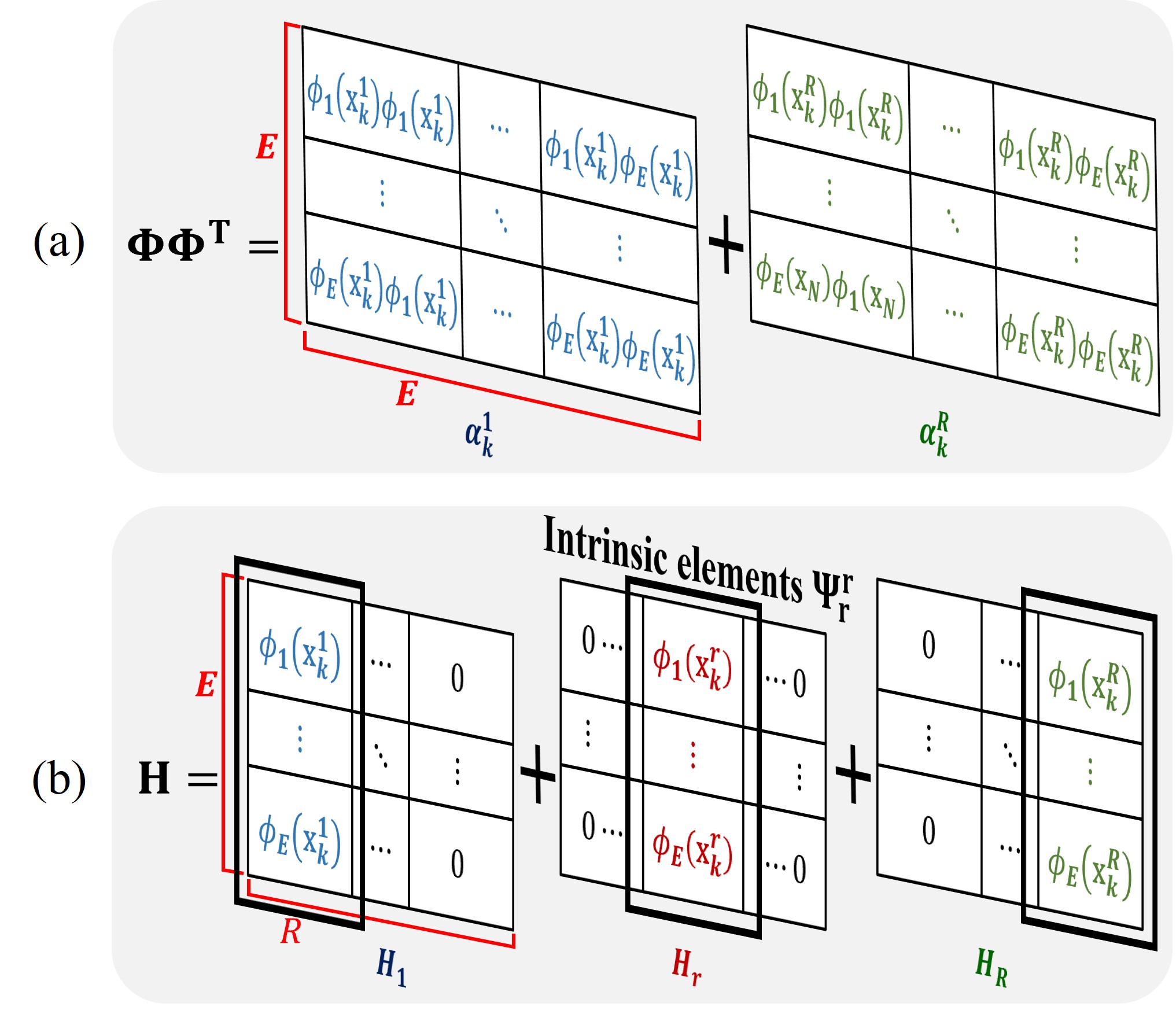}
	\caption{The message structure for communication. (a) The MADGP and (b) the K-DGP. Different colors are used to indicate the data from different sensors.}
	\label{phi_compare}
\end{figure}

The centralized information and the message required to communicate through the consensus protocol in both the MADGP and K-DGP, are illustrated in \figref{phi_compare}.
In the MADGP~(\figref{phi_compare}(a)), the message size is $\mathcal{O}(E^2)$, whereas in the K-DGP (\figref{phi_compare}(b)), it is $\mathcal{O}(E\cdot R)$. With a fixed number of wireless sensors, the message size in the MADGP increases quadratically as the number of nonlinear functions~$E$ grows for accurate kernel approximation, while the K-DGP shows a linear increase. Furthermore, when $E \gg R$, the computational complexity $\mathcal{O}(R^3)$ of the K-DGP associated with matrix inversion \eqref{KF_S_mul} is notably less than that of the MADGP $\mathcal{O}(E^3)$. 
Therefore, considering both communication and computation efficiency, the proposed approach exhibits clear advantages.

In real-world problems, it is highly likely that unknown scalar fields will change over time due to various factors such as convection and moving obstacles. The classic GP can estimate these dynamic environments using spatio-temporal kernels to model spatial and temporal correlations in the data, but they still encounter scalability issues. The MADGP could offer a solution by approximating spatio-temporal kernels for dynamic scalar field estimation. However, deriving the analytic form of the eigenfunctions of such kernels is a complex task (if not impossible), where only a few studies explore this area.

In the recursive Bayesian estimation for a single agent, the Kalman prediction process can be incorporated before the Kalman update~\citep{solin2018modeling}. We build upon this approach in the K-DGP, in order to provide a scalable method for estimating dynamic environments, even in the absence of an analytic eigenfunction form of the spatio-temporal kernel functions. Note that, unlike stationary cases, the obtained data $D_{1:k}={\{(k_i, \textbf{X}_i, \textbf{y}_i)\}}_{i=1}^k$ should incorporate a temporal variable which indicates the time step when the data is measured.

The composability of GP kernels allows for creating complex spatio-temporal kernels by combining simpler spatial and temporal ones through addition or multiplication. This is reversible; we can decompose a spatio-temporal kernel back into its individual components. In the temporal domain, kernel hyperparameters control the influence of past observations on future predictions, a concept known as the Ornstein-Ulenbeck process~\citep{rasmussen2004gaussian}:
\vspace*{-1.5mm}
\setlength{\arraycolsep}{0.0em}
\begin{eqnarray}
k_{temp}(k,k')=\exp{\left(-{\lvert k-k' \rvert}/{l_{k}}\right)}, \label{temp_GP}
\end{eqnarray}
\setlength{\arraycolsep}{5pt}
\vspace*{-6mm}

\noindent
where $l_k$ represents a temporal hyperparameter that determines the number of time steps, impacting the correlation among temporal variables. \cite{hartikainen2010kalman} derived the prediction step of the Kalman filter concerning the temporal GP kernel \eqref{temp_GP} as:
\vspace*{-2mm}
\setlength{\arraycolsep}{0.0em}
\begin{eqnarray}
\textbf{A}_k &=& \exp\left(-(\varDelta k)/{l_k}\right) \textbf{I}, \label{A}\\
\textbf{Q}_k &=& [1 - \exp\left(-2(\varDelta k)/{l_k}\right)]\textbf{I}, \\
\overline{\textbf{m}}_k &=& \textbf{A}_{k-1} \textbf{m}_{k-1}, \label{pred_m}\\
\overline{\textbf{P}}_k &=& \textbf{A}_{k-1} \textbf{P}_{k-1}\textbf{A}_{k-1}^\mathrm{T}+\textbf{Q}_{k-1}, \label{pred_P}
\end{eqnarray}
\vspace{-5mm}
\setlength{\arraycolsep}{5pt}

\noindent where $\varDelta k=k_{i+1} - k_i$. With this Kalman prediction step, $\textbf{m}_{k-1}$ and $\textbf{P}_{k-1}$ in  \eqref{KF_S_mul}$\sim$\eqref{KF_P_mul} of the Kalman update step should be replaced by \eqref{pred_m}$\sim$\eqref{pred_P}. 
Instead of using the spatio-temporal kernel, we use the Kalman prediction and update to deal with spatial and temporal correlation in the data separately.

Although it might be perceived as a mere extension of the K-GP algorithm's capabilities in dynamic scalar fields, our introduction of the K-DGP algorithm does not represent just a trivial advancement; it significantly enhances the feasibility of employing multiple sensors for distributed dynamic scalar field estimation, a task previously deemed highly complex.


\subsection{Dual-extrema consensus protocol}

In the K-DGP, we might employ the average consensus algorithm to converge each sensor's measurement matrix $\mathbf{H}_r$ to $\mathbf{H}$. However, unlike the message $\boldsymbol{\Phi}\boldsymbol{\Phi}^\mathrm{T}$ in the MADGP, the $\mathbf{H}_r$ contains a certain vector $\Psi^r_r$ (termed as intrinsic elements as shown in \figref{phi_compare}(b)) which needs to be maintained. 
These intrinsic elements are included in the $r$-th column which holds the only valid value for consensus and corresponds exactly to the $r$-th column of a centralized matrix. The average consensus protocol \eqref{avg2}, not accounting for this feature, can cause the intrinsic elements to be influenced by other sensor messages during the consensus process, potentially leading to a biased convergence. To address this issue, we propose a dual-extrema consensus protocol that preserves the intrinsic elements by concurrently taking the maximum and minimum values of the communicated matrix.
The dual-extrema consensus protocol is represented by using max-plus algebra~\citep{cohen1999max}:
\setlength{\arraycolsep}{0pt}
\begin{eqnarray}
\mathbf{H}_r(t+1)\:&=&\:\mathrm{Q}^+_r(t) + \mathrm{Q}^-_r(t), \label{privec 2}\\
\mathrm{Q}^+_r(t)\: &=&\: \underset{j\in\{r,\:\mathcal{N}_r\}}{\max}{\{\mathbf{H}_j(t)|\bar{\mathbf{E}}\}}, \label{positive}\\
\mathrm{Q}^-_r(t) \:&=& \:-\left[\underset{j\in\{r,\:\mathcal{N}_r\}}{\max}{\{-\mathbf{H}_j(t)|\bar{\mathbf{E}}\}}\right]. \label{negative}
\end{eqnarray}\noindent
where $t$ represents a consensus iteration, different from a sensing time step $k$, and $\bar{\mathbf{E}}$ is a zero matrix commonly used in max-plus algebra. The $\max\{\cdot \}$ operation in \eqref{positive}$\sim$\eqref{negative} adheres to the principles of max-algebra, as detailed in~\ref{apx max-pls alg}. The dual-extrema consensus protocol ensures that the intrinsic elements in $\Psi_r$ are not influenced by others, as it filters the values in the message into a non-negative matrix \eqref{positive} and a non-positive matrix \eqref{negative}. Consequently, each wireless sensor is able to achieve a consensus of $\mathbf{H}_r$ that is consistent with the centralized matrix $\mathbf{H}$. Compared with the average consensus protocol, the proposed protocol not only shows a faster convergence rate but also obviates the need for the consensus coefficient $\gamma$ in \eqref{avg2}. Furthermore, we have proven that the proposed protocol converges to the centralized matrix as provided in~\ref{apx proof}.

\SetKwComment{Comment}{/* }{ */}

\subsection{The flow of the K-DGP with dual-extrema consensus}
\setlength{\textfloatsep}{2pt}
\begin{algorithm}[t]
\caption{The proposed K-DGP with dual-extrema consensus for sensor $r$}\label{alg1}
	\KwRequire{$R$, $E$, $K_{max}$, $T_{max}$, $\theta_{th}$, \textit{Flag}, $\mathbf{m}_0$, $\mathbf{P}_0$}
	
	\For {$k=1,2,\ldots,K_{max}$}{
		$y_k^r = f(\textbf{x}^r_k) + \epsilon$\;  
		$\mathbf{H}_r(0) =
		\begin{bmatrix}
			\Psi_r^1(0) & \cdots & \Psi_r^R(0)
		\end{bmatrix}$\; 
		$t = 0,\; \Theta=\infty$\;
		\While{$t\leq T_{max}$ and $\Theta \geq \theta_{th}$}{
			$\mathbf{H}_r\left(t+1 \right) = \mathbf{Q}^+(t) + \mathbf{Q}^-(t)$\Comment*[r]{construct with $\textbf{y}_k$}
			$\Theta = \textbf{RMSE}(\mathbf{H}_r(t+1), \mathbf{H}_r(t))$\;
			$t = t+1$\;       
		}
		\eIf{Flag = True}{
			$\overline{\mathbf{m}}_k, \overline{\mathbf{P}}_k \gets$ \textbf{Prediction}$\left(\mathbf{m}_{k-1}, \mathbf{P}_{k-1} \right)$\;
			$\mathbf{m}_k, \mathbf{P}_k \gets$ \textbf{Update}$\left( \overline{\mathbf{m}}_k, \overline{\mathbf{P}}_k \right)$\;
		}
		{$\mathbf{m}_k, \mathbf{P}_k \gets$ \textbf{Update}$\left( \mathbf{m}_{k-1}, \mathbf{P}_{k-1} \right)$\;}    }
\end{algorithm}

To facilitate the understanding of the workflow of the K-DGP with dual-extrema consensus, we break down its operation for each sensor, as presented in Algorithm 1. In the proposed K-DGP algorithm with dual-extrema consensus, key parameters are $R$ for the number of sensors, $E$ for the number of eigenfunctions, $K_{max}$ for maximum sensing time, $T_{max}$ for the maximum number of iterations for convergence, $\theta_{th}$ as the threshold for convergence, and \textit{Flag} to indicate whether the dynamic field is considered or not. At each sensing step $k$, each sensor acquires measurements (line 2 of Algorithm~1). Using the measurements, each sensor constructs a matrix $\mathbf{H}_r$
to be shared (line 3 of Algorithm 1). Each sensor then exchanges this constructed matrix with its neighbors. With both received matrices from the neighbors and its own matrix, each sensor performs dual-extrema consensus update (line 6 of Algorithm 1) of \eqref{privec 2}$\sim$\eqref{negative}. The convergence of the matrix is evaluated by taking the root mean square error (RMSE) between the elements of previous $\mathbf{H}(t)$ and current $\mathbf{H}(t+1)$ matrices (line 7 of Algorithm~1). If the convergence is not achieved, the dual-extrema consensus iteration is repeated up to the predefined maximum number of iterations (lines 5-9 of Algorithm 1). For dynamic fields, the algorithm undergoes both KF prediction and update phases; otherwise, only the KF update phase is applied to obtain $\mathbf{m}_k$ and $\mathbf{P}_k$ (lines~10-15 in Algorithm 1).

\section{Numerical Simulations} \label{sec4}

In this section, we present a series of numerical simulations that illustrate the performance and benefits of the proposed method. The selected scenarios are intended to emphasize the unique features of the consensus protocol and the capabilities of the proposed K-DGP approach under various environments. 
\subsection{Consensus simulations under various network conditions}\label{sec4.A}
\begin{figure}[!t]
	\centering
	\includegraphics[width=\linewidth]{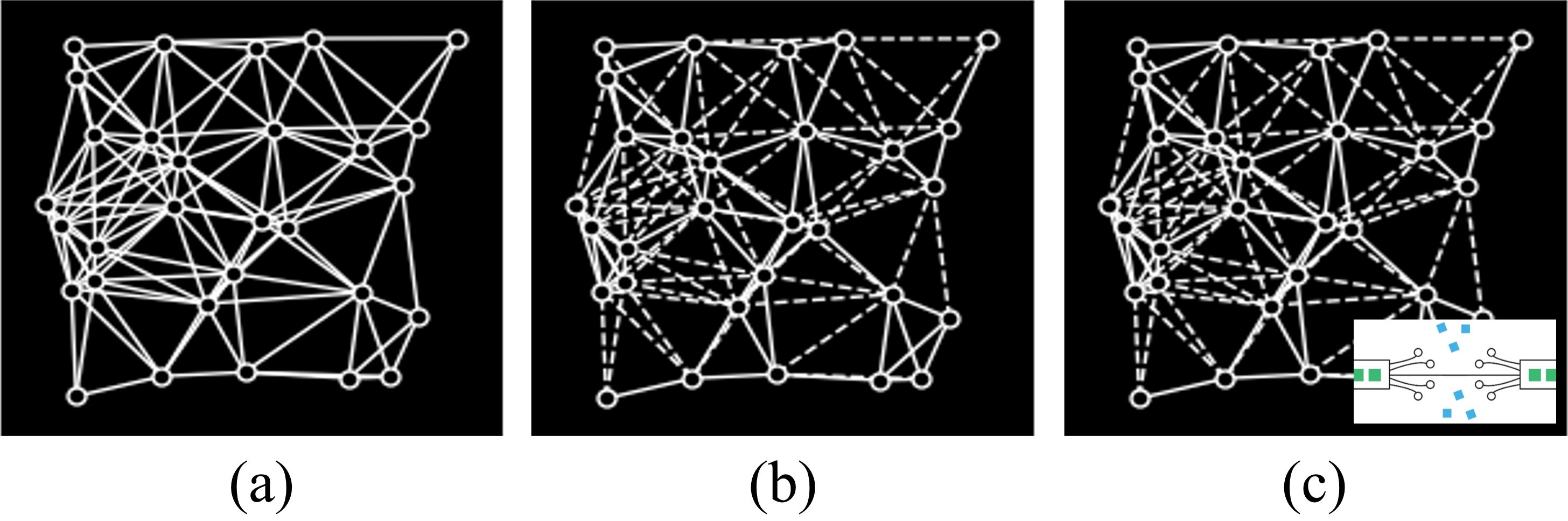}
	\caption{Network topology examples under three network conditions with 30 sensors. (a) Synchronized mesh network, (b) asynchronized mesh network where dashed lines indicate the links with the successful transmission probability of $p=0.3$, and (c) asynchronized mesh network with packet loss simulated by dropping rows from the row $i$ to the last row of the matrix at every iteration where the row $i$ is randomly selected from the uniform distribution.}
	\vspace{5.mm}
	\label{network conditions}
\end{figure}
\begin{figure}[!t]
	\centering
	\includegraphics[width=\linewidth]{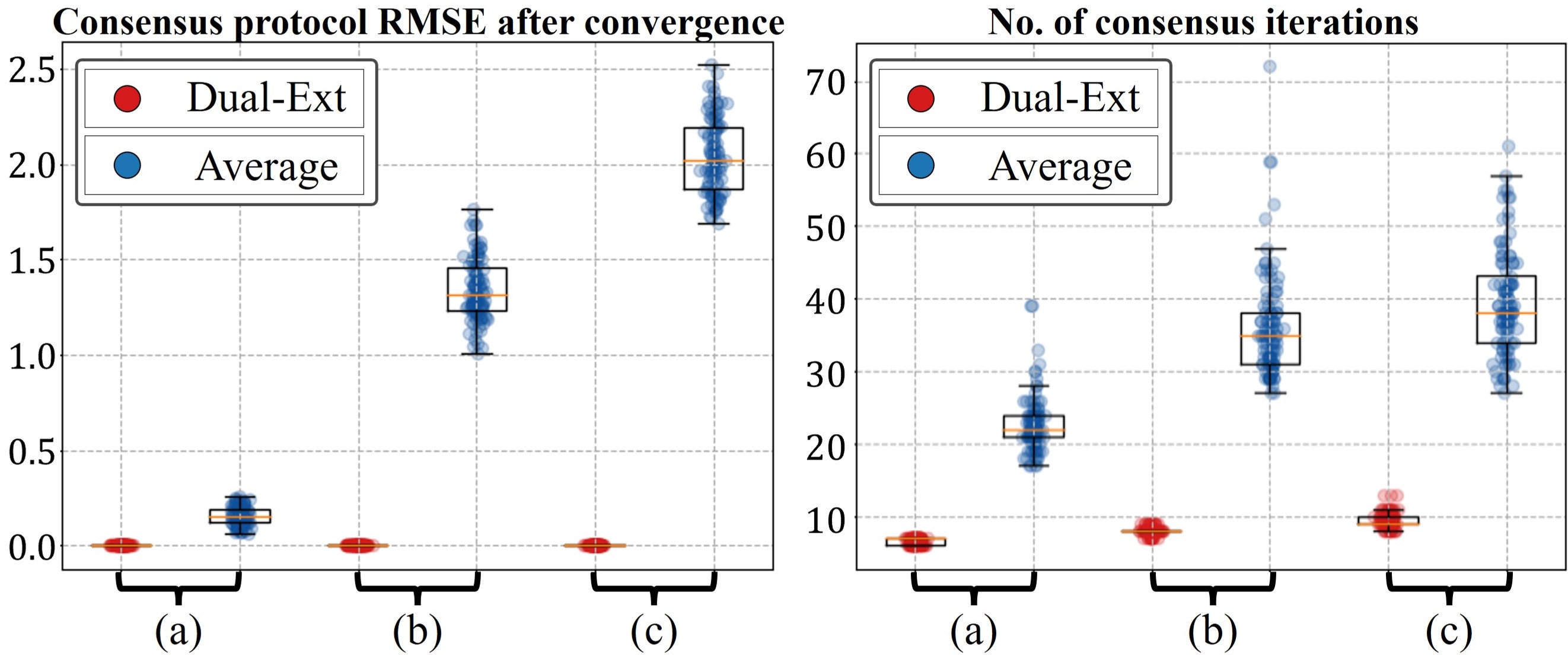}
	\caption{Comparison of RMSE and the number of iterations for convergence between average and dual-extrema consensus in distributed networks. (a), (b), and (c) are the represented networks in \figref{network conditions}.}
	\vspace{5.mm}
	\label{network simulations}
\end{figure}
To compare the performance of the average~\eqref{avg2} and dual-extrema~\eqref{privec 2}$\sim$\eqref{negative} consensus protocols, we conducted consensus simulations under three different network conditions as shown in \figref{network conditions}. In these simulations, we used randomly generated network topologies and matrices which of each column consists of the intrinsic elements. For the comparison with the average consensus, the matrices are scaled by the number of sensors. 

We evaluated the protocols in terms of accuracy using RMSE comparisons with a centralized matrix and in terms of the number of consensus iterations. Consensus criteria is defined as the point where the algorithm's matrix remains unchanged over three consecutive iterations within a 0.01 threshold. 
Each network condition is simulated with 100 trials and then the results are averaged.
As shown in \figref{network simulations} (left), the  consensus fully converges to a centralized matrix under the complex network conditions.
In comparison, as the network conditions get worse, RMSE values of the average consensus method increase.
In \figref{network simulations} (right), the dual-extrema consensus not only reaches consensus more rapidly than the average consensus but also keeps data variance low. This low variance leads to more reliable and consistent results, further enhancing the algorithm's practical applicability. In summary, this comparative analysis demonstrates that the dual-extrema consensus protocol exhibits superior performance to the average consensus within the K-DGP framework.

\subsection{Stationary Field Estimation} \label{station sim}
The stationary scalar fields are first used to validate the proposed K-DGP approach using the squared exponential kernel with hyperparameters $( \sigma_s = 4$ and $l = 0.05)$. In the same way as the K-GP, we employed the Hilbert space approximation to represent the eigenfunctions and eigenvalues as~\citep{solin2014hilbert}:
\setlength{\arraycolsep}{0.0em}
\begin{eqnarray}
&&\phi_{e}(\textbf{x})=\prod_{m=1}^{2} \frac{1}{\sqrt{L}} \sin\left(\frac{\pi j_{e,m} (x_m + L) }{2L}\right), \label{eigenfunction} \\
&&\lambda_{e}=\sum_{m=1}^{2} \left(\frac{\pi j_{e,m}}{2 L}\right)^2,
\end{eqnarray}
\setlength{\arraycolsep}{0pt}\noindent
where $\textbf{x}=[x_1,\:x_2]^\mathrm{T}$ and $\textbf{j}_e=[{j_{e,1}},\:{j_{e,2}}]$ represent selected pairs of integers $(1,2,\dots,E)$, (e.g., $[1,1],[1,2],\dots,[E,1],\dots$). The spectral density \eqref{lambdaS} of the squared exponential kernel \eqref{sekern} is then represented as~\citep{rasmussen2004gaussian}:
\setlength{\arraycolsep}{0.0em}
\begin{equation}
S(\lambda_e)=\sigma_s^2 {\left(2\pi l\right)}^{3/2} \exp\left(-l^2 \lambda_e/2\right).
\end{equation}
\setlength{\arraycolsep}{5pt}\noindent	

\begin{figure}[p]
\centering
\includegraphics[width=\linewidth]{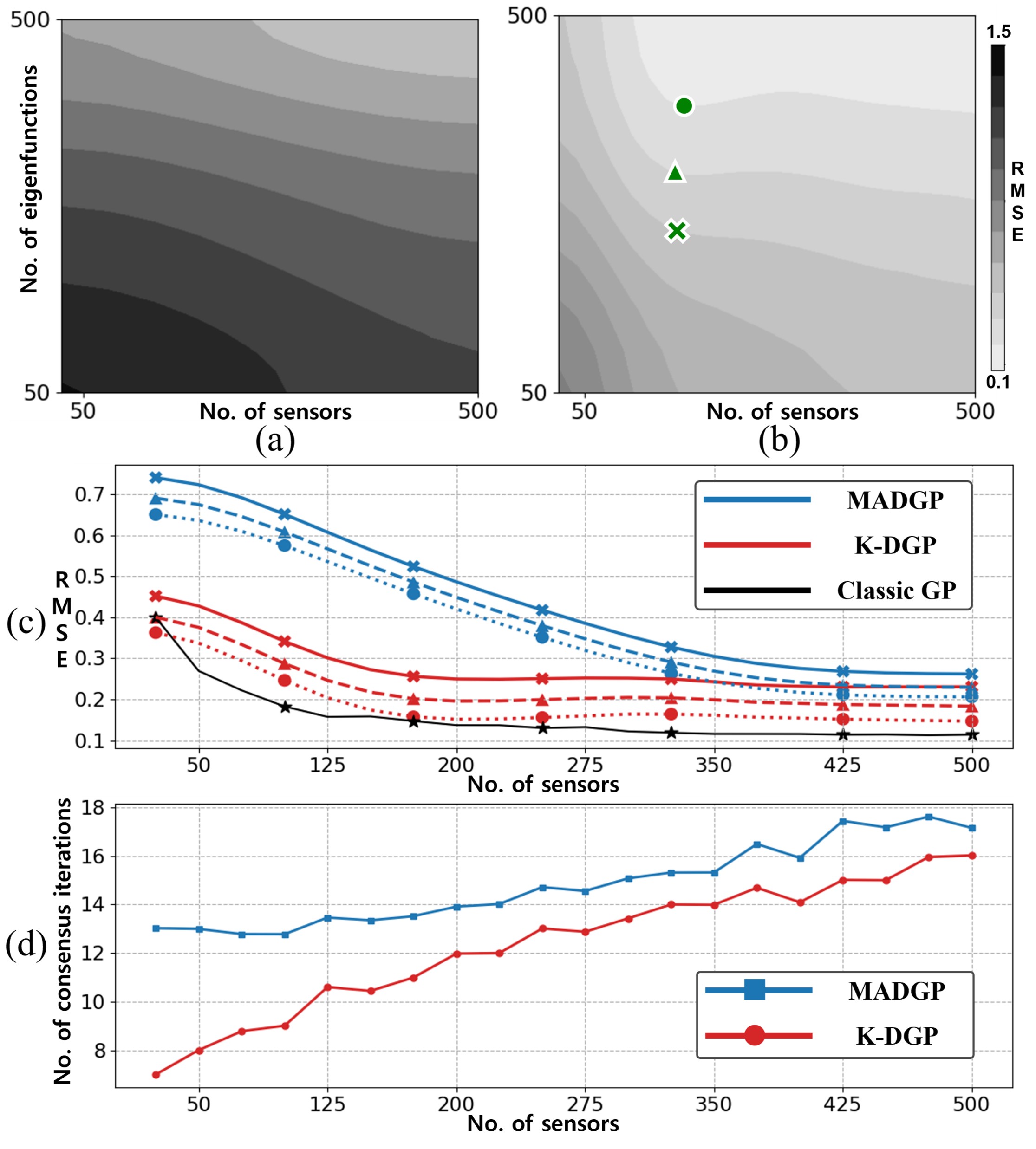}
\caption{Performance comparison between two approaches. (a) RMSE of MADGP, (b) RMSE of K-DGP (circle: $E=400$, triangle: $E=325$, and cross: $E=250$), (c) comparison with the centralized GP for cases using the same number of eigenfunctions marked in (b), and (d) the number of consensus iterations when $E=400$.}
\label{DGP rmse compare}
\end{figure}

The number of sensors ranged from $25$ to $500$, and the number of basis functions varied from $50$ to $500$. Each scenario was subjected to 100 trials, and the scalar fields were randomly sampled from a Gaussian kernel per each trial. 
The performance of the algorithms is presented in \figref{DGP rmse compare}, evaluated by RMSE between the algorithms and the true scalar field, consensus iterations, and the computation/communication complexities.
\begin{figure}[!t]
\centering
\includegraphics[width=\linewidth]{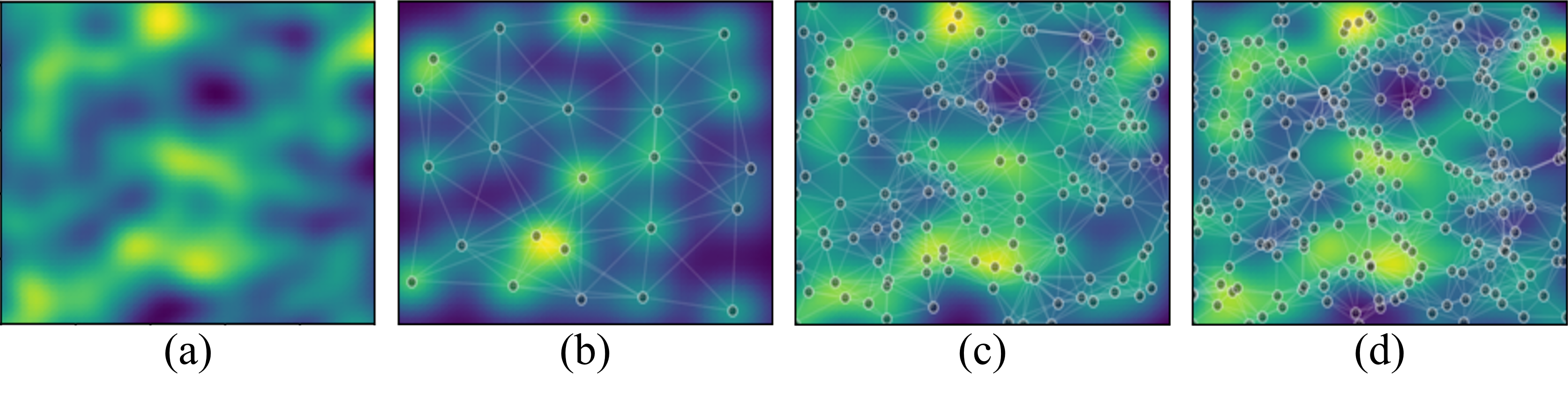}
\caption{Scalar field estimation results of K-DGP when E=400. (a) True scalar field, (b) $R=25$, (c) $R=150$, and (d) $R=250$.}
\vspace{5.mm}
\label{Sensor estimation results}
\end{figure}		

In a fixed-size domain, it is observed that a threshold exists beyond which adding more sensors does not significantly reduce the error, marked by three saturation points in \figref{DGP rmse compare}(b). To illustrate this error reduction behavior, the estimated scalar field results obtained using varying numbers of sensors are visualized in \figref{Sensor estimation results}; the estimation result does not change much from $R=150$ to $R=250$.
Therefore, optimizing performance is more effectively achieved by increasing the number of eigenfunctions rather than by adding more sensors especially for K-DGP. In situations where $E \gg R$, the proposed K-DGP approach outperforms MADGP in terms of computational complexity $\mathcal{O}(R^3)$ and communication message size $\mathcal{O}(E\cdot R)$, compared to MADGP's computational complexity $\mathcal{O}(E^3)$ and communication message size of $\mathcal{O}(E^2)$.

With the aforementioned sample saturation points, RMSE of both algorithms against the classic GP (i.e., centralized result) is presented in \figref{DGP rmse compare}(c).
Consistent with \figref{kernel_approx}, increasing eigenfunctions narrows the RMSE gap between both algorithms and the classic GP.
Notably, although employing a larger number of sensors makes the RMSE difference between MADGP and K-DGP smaller, K-DGP consistently shows lower RMSE values compared to MADGP when the same number of eigenfunctions are used.


The comparison of the number of consensus iterations between MADGP and K-DGP is presented in \figref{DGP rmse compare}(d). The consensus criteria is defined the same as in \secref{sec4.A}. As the number of sensors in the network increases, the size of the network correspondingly expands, thereby requiring a greater number of consensus iterations. Although the performance gap between the two algorithms narrows as the network size expands, our proposed algorithm, K-DGP, reaches consensus more rapidly than MADGP.


These comparative analysis highlight the advantages of using the K-DGP with the dual-extrema consensus over the MADGP with the average consensus. With its superior estimation accuracy and reduced communication complexity, the K-DGP presents a promising option for various applications, particularly when dealing with complex scalar fields. Furthermore, the faster convergence of the K-DGP implies that it is a more time-efficient and computationally effective solution, which is essential for real-world scenarios where resources and time are often limited.

\subsection{Spatio-Temporal Field Estimation}

In order to validate the feasibility and performance of the proposed algorithm in dynamic scalar fields, we constructed a scenario based on the convection-diffusion equation, which is widely utilized in environmental sciences \citep{bejan2013convection}. The equation is represented as:
\setlength{\arraycolsep}{0.0em}
\begin{equation}
\frac{\partial f(\textbf{x},t)}{\partial t} - \nabla \cdot (v(\textbf{x},t)f(\textbf{x},t)) = \nabla \cdot (D\nabla f(\textbf{x},t)) + C(\textbf{x}). \label{dynamic field}
\end{equation}
\setlength{\arraycolsep}{5pt}\noindent
In this equation, the diffusivity function is denoted by $D(\textbf{x})$:
\setlength{\arraycolsep}{5pt}
\begin{eqnarray}
D(\textbf{x}) = 0.005(x_1^2 + x_2^2) + 0.01 x_1 x_2 + 0.02.
\end{eqnarray}
\setlength{\arraycolsep}{5pt}\noindent
The fluid's velocity, which could be wind or water, is represented by $v(\textbf{x},t)$:
\setlength{\arraycolsep}{5pt}
\begin{eqnarray}
v(\textbf{x},t)=
\begin{bmatrix}
2(x_1 + x_2 -t), & x_2 - x_1 + t
\end{bmatrix}^\mathrm{T}.
\end{eqnarray}
\setlength{\arraycolsep}{5pt}\noindent
The scalar source function $C(\textbf{x})$ characterizes the quantity $f(\textbf{x},t)$ in the following manner:
\setlength{\arraycolsep}{5pt}
\begin{eqnarray}
C(\textbf{x})=\exp\left(- \frac{{(x_1 - c_1)}^2 + {(x_2 - c_2)^2}}{0.007} \right), \label{dynamic coeff}
\end{eqnarray}
\setlength{\arraycolsep}{5pt}\noindent
where $\textbf{x}=[x_1, x_2]^\mathrm{T}$ and $\textbf{c}=[c_1, c_2]^\mathrm{T}$ represent the source location of the convection-diffusion model. In our scenario, the source is located at $\textbf{c}=[6,6]$. Since finding analytical solutions to partial differential equations (PDE) is typically a challenging task, we employed the finite difference method for obtaining numerical solutions to the PDE. Utilizing these numerical solutions, we then conducted the corresponding simulations in order to model the dynamic scalar field.
\begin{figure}[t]
\centering
\includegraphics[width=\linewidth]{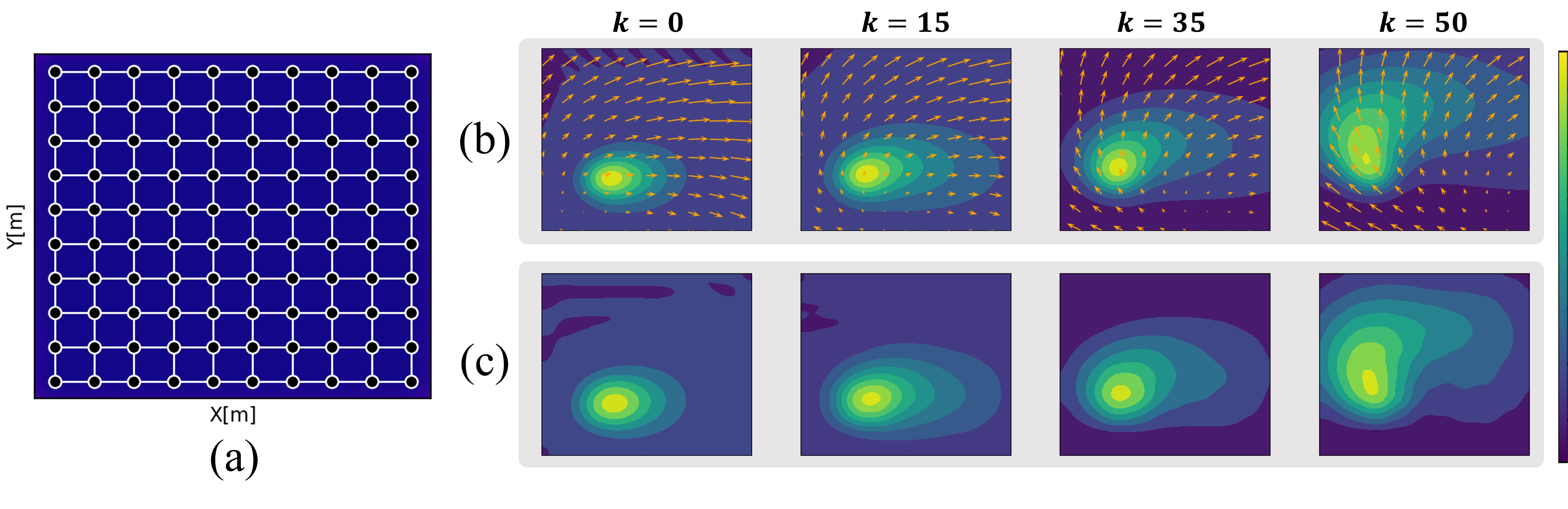}
\caption{The results of a dynamic scalar field estimation using K-DGP with dual-extrema consensus protocol. (a) Black dots are deployed WSNs and white edges represent the network topology, (b) true dynamic scalar field with the orange-colored wind vector arrows and (c) estimated dynamic scalar field.}
\label{scenarios_dynamic}
\end{figure}	
We use $R=100$ wireless sensors as shown in \figref{scenarios_dynamic}(a) and set $E=300$ for accurate estimation. Due to the larger network size, the consensus iteration proceeded for $15$ iterations at each time step $k$, and $\varDelta k = 25$ and $l_k=3600$ in \eqref{A}. Other configurations are the same as those used in \secref{station sim}. 

The simulation results are shown in \figref{scenarios_dynamic}. At each time step $k$, the true dynamic scalar field slightly changes according to \eqref{dynamic field}$\sim$\eqref{dynamic coeff}, as illustrated in \figref{scenarios_dynamic}(b). 
The proposed algorithm can effectively estimate the dynamic unknown scalar field, as demonstrated in \figref{scenarios_dynamic}(c) without using the wind vector. This capability renders the algorithm more suitable for real-world applications compared to the MADGP, providing a robust and reliable solution for dynamic scenarios. It is worthwhile mentioning that DGP studies for dynamic fields are rarely reported.

\section{Conclusions and Future Work} \label{sec5}

For online estimation of unknown scalar fields, we have presented a Kalman filter-based DGP (K-DGP) framework with a consensus protocol which is tailored to the K-DGP. Although the kernel-based GP is widely used for scalar field estimation, their centralized characteristic poses challenges for handling large amounts of data from wireless sensor networks (WSNs). Although nonlinear function-based GP methods such as the MADGP could be used, they are still limited by the size of $E$ (i.e., the number of eigenfunctions).
As another type of nonlinear function-based GP, the proposed K-DGP overcomes these limitations by being independent of the data size and less dependent on the size of $E$. 
The number of eigenfunctions influences both the computational complexity and the message size required to communicate. As the number of wireless sensors $R$ is typically less than $E$, the K-DGP reduces computational complexity to $\mathcal{O}(R^3)$ and communication complexity to $\mathcal{O}(E)$, in comparison with $\mathcal{O}(E^3)$ and $\mathcal{O}(E^2)$ of the MADGP. We also introduced the dual-extrema consensus protocol, which preserves the intrinsic elements in the communicated message. This protocol enables faster convergence to the centralized matrix (i.e., unbiased information), as demonstrated in an illustrative consensus simulation. Given that unknown scalar fields can be represented by any information distribution measured with the Gaussian noise, we validated the proposed approach against the MADGP using a randomly generated complex scalar field. The results showed that the K-DGP with the dual-extrema protocol outperforms the MADGP in terms of both estimation accuracy and communication complexity. Moreover, we demonstrated the applicability of the K-DGP in a dynamic environment with its ability to consider temporal variations in the data by sequentially incorporating new data.

While static sensor networks with a sufficient number of sensors can estimate a wide environment, a small number of mobile sensors could provide a more effective solution for online estimation. To achieve this, mobile sensors need to choose actions that maximize the information gain in a distributed manner, which is referred to informative path planning.
For instance, GP variance or mutual information of maximum field values \citep{SEO2023119459} could be used as a possible metric; this will be performed as the future work.

\section{Acknowledgments}

This research was supported by the Defense Challengeable Future Technology Program of Agency for Defense Development, Republic of Korea (UC200016RD).

\appendix
\section{Max-Plus Algebra} \label{apx max-pls alg}
In developing the proposed dual-extrema consensus protocol, we utilized the max-plus algebra, a mathematical structure with diverse applications including graph theory~\citep{cohen1999max}. Max-plus algebra differs from conventional algebra by replacing traditional arithmetic operations such as addition and multiplication with operations of maximum and addition, respectively. In this appendix, we briefly explain the concept of the max-plus algebra, borrowed from~\citep{nejad2009max}.
\begin{definition}
	In the max-plus algebra, the set of real numbers $\mathbb{R}$ is extended by including $-\infty$, forming the set $\mathbb{R}_{\max}:= \mathbb{R}\cup\{-\infty\}$. This expanded set is equipped with two basic operations:
	\begin{eqnarray}
		a \oplus b &=& \max(a,b), \\
		a \otimes b &=& a + b.
	\end{eqnarray}
\end{definition}
The additive identity element is ${\varepsilon}=-\infty$ (i.e., $a \oplus \varepsilon=a$), and the multiplicative identity element is $e=0$ (i.e., $a \otimes e= a$). The max-plus algebra retains some properties from classical algebra, such as associativity, commutativity, and distributivity. 

For matrices $A, B\in \mathbb{R}_{\max}^{M\times N}$, the addition of the max-plus algebra are extended as follows:
\begin{eqnarray}
	A \oplus B = \left[a_{ij} \oplus b_{ij}\right]_{1\le i \le M,\; 1\le j \le N}. \notag
\end{eqnarray}
The max-plus multiplication of two matrices $A\in \mathbb{R}_{\max}^{M\times N}$ and $B\in \mathbb{R}_{\max}^{N \times Q}$ is represented as:
\setlength{\arraycolsep}{0pt}
\begin{eqnarray}
	A \otimes B &=& \left[\bigoplus_{n=1}^N \left( a_{in} \otimes b_{nj} \right)\right]_{1\le i \le M,\; 1\le j \le Q}, \notag\\
	&=&\left[ \underset{n\in\{1,\dots,N\}}{\max}\left( a_{in} + b_{nj} \right)\right]_{1\le i \le M,\; 1\le j \le Q} \in \mathbb{R}^{M\times Q}_{\max}. \label{max_mat_mul}
\end{eqnarray}
\setlength{\arraycolsep}{5pt}\noindent

In max-plus algebra, a matrix whose whose elements are all zero, denoted by $e$, is referred to as $\bar{\mathbf{E}}$:
\begin{eqnarray}
	\bar{\mathbf{E}}=
	\begin{bmatrix}
		e& \cdots & e\\
		\vdots& \ddots & \vdots \\
		e& \cdots & e
	\end{bmatrix}. \label{zero matrix}
\end{eqnarray}

\begin{figure}[t]
	\centering
	\includegraphics[width=\linewidth]{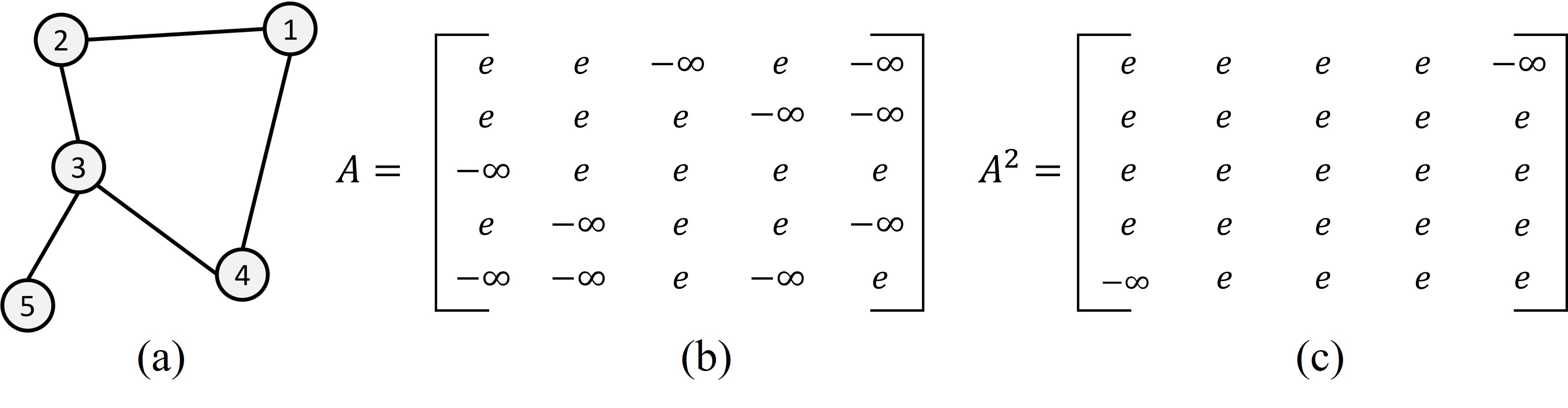}
	\caption{(a) Illustrative network topology, (b) adjacency matrix $A$ and (c) the power of the adjacency matrix $A$.}
	\vspace{5.mm}
	\label{topology}
\end{figure}

In the communication network modeled by an undirected graph $\mathcal{G}(\mathcal{V},\mathcal{E})$, its structure is encapsulated in an adjacency matrix $A$, as illustrated in \figref{topology}. Each element $a_{ij}$ of matrix $A$ indicates whether node $i$ is capable of receiving information from node $j$. Here, $a_{ij}=e$ represents a direct information pathway between the two nodes, while $a_{ij}=\epsilon$ denotes the absence of such a pathway. For clarify in representation, $-\infty$ is employed to represent $\epsilon$ in \figref{topology}(b).
The $t$-th power of an adjacency matrix $A$, denoted as $A^t$, is utilized to analyze the graph's connectivity over $t$ iterations using max-plus multiplication \eqref{max_mat_mul}:
\begin{eqnarray}
	A^{t} := \underbrace{A \otimes A \otimes \cdots \otimes A}_{(t-1)\text{-times multiplication}}, \quad t \geq 1.
\end{eqnarray}
This metric articulates the potential paths where information from node $i$ can reach node $j$ within $t$ steps. For instance, in the scenario depicted in \figref{topology}(c), the matrix $A^2$ demonstrates the network's two-step communication paths.

\section{The proof of convergence with the dual-extrema consensus protocol} \label{apx proof}
\begin{figure}[h]
	\centering
	\includegraphics[width=\linewidth]{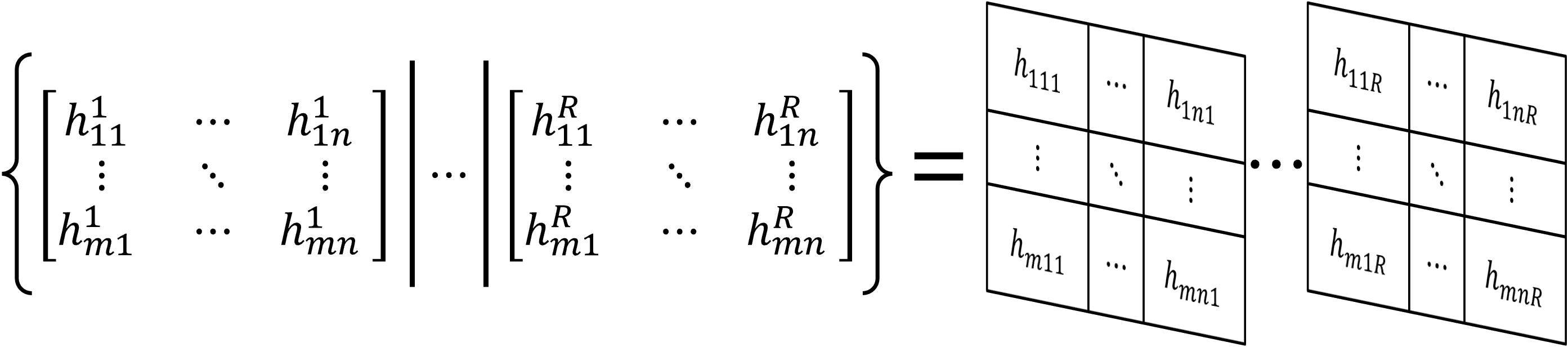}
	\caption{Example structure of a three-dimensional matrix array.}
	\vspace{5.mm}
	\label{three-dimensional matrix}
\end{figure}

In this appendix, we extend the existing proof of the convergence of the max-consensus protocol~\citep{nejad2009max} for the dual-extrema consensus protocol. Note that, the notations $\left[\;\right]$ and $\{\;\}$ are distinguished between two-dimensional and three-dimensional matrices, respectively.
\begin{definition}
	Consider a series of $M \times N$ two-dimensional matrices, denoted as $\mathbf{H}_1, \dots, \mathbf{H}_R$. A three-dimensional array matrix, represented as $\widehat{\mathbf{H}}$, is constructed by sequentially stacking these matrices. Specifically, this array is arranged such that:
	\begin{eqnarray}
		\widehat{\mathbf{H}} =\{\mathbf{H}_1|\dots|\mathbf{H}_R \}=\{h_{ijr}\}_{1\le i \le M,\; 1\le j\le N,\; 1\le r \le R}. \label{three dim mat}
	\end{eqnarray}
\end{definition}
Figure \ref{three-dimensional matrix} illustrates the three-dimensional matrix structure.

\begin{definition}
	A three-dimensional matrix $\widehat{\mathbf{H}}$ is transformed into a two-dimensional matrix by performing a max operation, represented as:
	\setlength{\arraycolsep}{0pt}
	\begin{eqnarray}
		\max{\widehat{\mathbf{H}}}&=&\max{\{\mathbf{H}_1|\dots|\mathbf{H}_R \}}
		= \underset{r \in \{1,\dots,R\}}{\max}\{\mathbf{H}_r\} , \\
		&=&\left[\bigoplus_{r=1}^R{h_{ijr}} \right]_{1\le i \le M, \; 1 \le j \le N }.
	\end{eqnarray}
	\setlength{\arraycolsep}{5pt}\noindent
\end{definition}
This transformation involves selecting the maximum value for each element across the third dimension (depth) of the array.


Before we proceed with the proof of the dual extrema consensus protocol, it is essential to establish a key assumption. This is due to the fact that in our framework, each sensor $r$ communicates a matrix where specific intrinsic elements are contained in column $\Psi_r$, while the other columns consist of zero vectors.
In the three-dimensional matrix $\widehat{\mathbf{H}}$, which consists of the two-dimensional matrices $\mathbf{H}_1,\dots,\mathbf{H}_R$, our assumption is as follows:
\begin{assumption}
	For any element at position $(i,j)$ in the two-dimensional matrices, all corresponding elements along the third dimension—located at $(i,j,1),(i,j,2),$ $\dots,(i,j,R)$ in $\widehat{\mathbf{H}}$—are either identical or zero. \label{assumption1}
\end{assumption}
To further illustrate this assumption, consider a simplified example where $R=3$:
\begin{eqnarray}
	\mathrm{H}_1=
	\begin{bmatrix}
		1 & 0 \\
		-7 & 0 
	\end{bmatrix},\;
	\mathrm{H}_2=
	\begin{bmatrix}
		0 & 4 \\
		-7 & 0
	\end{bmatrix},\;
	\mathrm{H}_3=
	\begin{bmatrix}
		0 & 4\\
		-7 & 2
	\end{bmatrix}. \label{assumption matrix example}
\end{eqnarray}
This assumption paves the way for \textbf{Theorem~\ref{theorem1}}, where we can see how matrices with such structures assist in the formation of a non-negative matrix $\mathbf{Q}^+$.

\begin{theorem}
	Under the \textbf{Assumption~\ref{assumption1}}, the max operation of $\{\widehat{\mathbf{H}}|\bar{\mathbf{E}}\}$ results in a non-negative matrix $\mathbf{Q}^+$: 
	\setlength{\arraycolsep}{0pt}
	\begin{eqnarray}
		\mathbf{Q}^+&=&\max{\{\widehat{\mathbf{H}}|\bar{\mathbf{E}}\}}=\max{\{\mathbf{H}_1|\dots|\mathbf{H}_R|\bar{\mathbf{E}} \}},\notag \\
		&=&\underset{r\in \{1,\dots,R\}}{\max}\{\mathbf{H}_r|E\},\notag \\
		&=&\left[\bigoplus_{r=1}^{R+1}{h_{ijr}} \right]_{1\le i \le M, \; 1 \le j \le N },
	\end{eqnarray} \label{theorem1}
\end{theorem}
where $h_{ij(R+1)} = e$.
\begin{proof}
	Let matrices $\widehat{\mathbf{H}}$ conform to \textbf{Assumption~\ref{assumption1}} and consider a zero matrix $\bar{\mathbf{E}}$ in \eqref{zero matrix}, where each element is $e_{ij} (= e)$. For the operation $h_{ijr}\oplus e_{ij}$, by the definition of $\oplus$, it results in $h_{ijr}$ if $h_{ijr} > 0$ or $e_{ij}$ otherwise. Since both filtered $h_{ijr}$ and $e_{ij}$ are non-negative, each element in the resulting matrix from  $\{\widehat{\mathbf{H}}|\bar{\mathbf{E}}\}$ have non-negative value, leading to a non-negative matrix. 
\end{proof}\noindent
For matrices $\mathrm{H}_1, \mathrm{H}_2,$ and $\mathrm{H}_3$ (as exampled in \eqref{assumption matrix example}), we can obtain the non-negative matrix $\mathbf{Q}^+$ as follows:
\begin{eqnarray}
	\mathbf{Q}^+={\max}\{\mathbf{H}_1|\mathbf{H}_2|\mathbf{H}_3|\bar{\mathbf{E}}\}
	=
	\begin{bmatrix}
		1 & 4 \\
		0 & 2
	\end{bmatrix}. \notag
\end{eqnarray}
Following \textbf{Theorem~\ref{theorem1}}, we can similarly define a non-positive matrix $\mathbf{Q}^-$ as follows:
\setlength{\arraycolsep}{0pt}
\begin{eqnarray}
	\mathbf{Q}^-&=&-\left[\max{\{-\widehat{\mathbf{H}}|\bar{\mathbf{E}}\}}\right], \notag \\
	&=&-\left[\max\{-\mathbf{H}_1|-\mathbf{H}_2|-\mathbf{H}_3|\bar{\mathbf{E}}\} \right],\notag \\
	&=& 
	\begin{bmatrix}
		0 & 0 \\
		-7 & 0
	\end{bmatrix}. \notag
\end{eqnarray}

Next, we introduce the concept of dual-extrema consensus achievement, which considers the neighbor set of node $r$, $\mathcal{N}_r=\{j | \mathcal{E}(r,j) = e\}$ and the set of entire nodes $\mathcal{N}$. 
\begin{definition}
	Given an initial matrix of information states as
	\begin{eqnarray}
		\widehat{\mathbf{H}}(t):=\left[\mathbf{H}_1(t)|\dots|\mathbf{H}_R(t)\right],
	\end{eqnarray} the dual-extrema consensus is said to be achieved at the consensus iteration $t$ if:
	\setlength{\arraycolsep}{0pt}
	\begin{eqnarray}
		\mathbf{H}_i(t) &=& \mathbf{H}_j(t),\label{A7} \\
		&=& \mathbf{Q}^+(0) + \mathbf{Q}^-(0),\quad \forall i,j \in \mathcal{N}. \label{A8}
	\end{eqnarray}
	where,
	\setlength{\arraycolsep}{0pt}
	\begin{eqnarray}
		\mathbf{Q}^+(0) &=& {\max}\{\mathbf{H}_1(0)|\dots|\mathbf{H}_R(0)|\bar{\mathbf{E}}\}, \\
		\mathbf{Q}^-(0) &=& -\left[{\max}\{-\mathbf{H}_1(0)|\dots|-\mathbf{H}_R(0)|\bar{\mathbf{E}}\}\right]. 
	\end{eqnarray} \label{def4}
\end{definition}
This definition implies that consensus is reached when all nodes $i,j \in \mathcal{N}$ have identical information states at iteration $t$, represented as the sum of the initial maximum and minimum extremal values from their respective node information states. In order to achieve the consensus as outlined in \textbf{Definition~\ref{def4}}, we consider the iterative update of the communicated matrix of each sensor $r$ as:
\begin{eqnarray}
	\mathbf{H}_r(t+1) = \mathbf{Q}^+_r(t) + \mathbf{Q}^-_r(t),\; \forall r \in \mathcal{N}, \label{H P N}
\end{eqnarray}
where,
\setlength{\arraycolsep}{0pt}
\begin{eqnarray}
	\mathbf{Q}^+_r(t) &=& \underset{j\in \{\mathcal{N}_r, r\}}{\max}\{\mathbf{H}_j(t)|\bar{\mathbf{E}}\}, \label{P_t}\\
	\mathbf{Q}^-_r(t) &=& -\left[\underset{j\in \{\mathcal{N}_r, r\}}{\max}\{-\mathbf{H}_j(t)|\bar{\mathbf{E}}\}\right].\label{N_t}
\end{eqnarray}

Following the establishment of the iterative update process in \eqref{H P N}$\sim$\eqref{N_t}, we now introduce the influence of the adjacency matrix $A$ into the consensus process of the matrix $\widehat{\mathbf{H}}(t)$, defining the max-plus multiplication operation between $A$ and $\widehat{\mathbf{H}}(t)$. 
\begin{definition}
	The max-plus multiplication of between the adjacency matrix $A\in \mathbb{R}^{R\times R}_{\max}$ and a three-dimensional matrix $\widehat{\mathbf{H}}(t)$ is defined as:
	\begin{eqnarray}
		A\otimes \widehat{\mathbf{H}}(t) = \{ \boldsymbol{a}_1 \otimes \widehat{\mathbf{H}}(t)|\boldsymbol{a}_2\otimes \widehat{\mathbf{H}}(t)|\dots|\boldsymbol{a}_R\otimes\widehat{\mathbf{H}}(t)\},
	\end{eqnarray}
	where,
	\setlength{\arraycolsep}{0pt}
	\begin{eqnarray}
		A&=&
		\begin{bmatrix}
			\boldsymbol{a}_1\quad&\cdots&\quad\boldsymbol{a}_R
		\end{bmatrix}^\mathrm{T}, \\
		\boldsymbol{a}_i \otimes \widehat{\mathbf{H}}(t)&=&\left[\bigoplus_{r=1}^R (a_{ir} \otimes h_{jkr}) \right]_{1\le j \le M,\; 1\le k \le N}, \label{neighbor max}\\
		&=&\left[\underset{r\in\{1,\dots,R\}}{\max}{(a_{ir}\otimes h_{jkr})}\right]_{1\le j \le M,\; 1\le k \le N}.
	\end{eqnarray}
	\setlength{\arraycolsep}{5pt}\noindent
\end{definition}    
An element in a row $\boldsymbol{a}_i$ is denoted as $e$ when it corresponds to a neighbor of node $i$; otherwise, it is represented by $-\infty$. Consequently, the operation $a_{ir} \otimes h_{jkr}$ yields $h_{jkr}$ if $a_{ir}=e$, and results in $-\infty$ if $a_{ir}=-\infty$. Therefore, Equation \eqref{neighbor max} represents the max operation on three-dimensional matrices involving neighboring elements.
Accordingly, the update for each node $r$ in \eqref{H P N}$\sim$\eqref{N_t} can be collectively encapsulated in the three dimensional matrix $\widehat{\mathbf{H}}(t+1)$ as:
\begin{eqnarray}
	\widehat{\mathbf{H}}(t+1)=A \otimes \{\widehat{\mathbf{H}}(t)|E\} - \left[A \otimes \{-\widehat{\mathbf{H}}(t)|E\}\right]. \label{H_t}
\end{eqnarray}
The recursive form of \eqref{H_t} is represented as:
\begin{eqnarray}
	\widehat{\mathbf{H}}(t) = A^t \otimes \{\widehat{\mathbf{H}}(0)|E\} - \left[A^t \otimes \{-\widehat{\mathbf{H}}(0)|E\}\right].
\end{eqnarray}

\begin{theorem}
	Let $A$ be an adjacency matrix of a graph $\mathcal{G}(\mathcal{V}, \mathcal{E})$. If there exists an integer $t_*$ such that $A^{t_*} =\bar{\mathbf{E}}\:$, then the dual-extrema consensus is achieved.
\end{theorem}

\begin{proof}
	$A^{t_*}=\bar{\mathbf{E}}$ implies,
	\begin{eqnarray}
		\widehat{\mathbf{H}}(t_*)= \bar{\mathbf{E}} \otimes \{\widehat{\mathbf{H}}(0)|\bar{\mathbf{E}}\} - \left[\bar{\mathbf{E}} \otimes\{-\widehat{\mathbf{H}}(0)|\bar{\mathbf{E}}\} \right], \label{H_t proof}
	\end{eqnarray}
	and the first term of the right-hand side is represented as:
	\begin{eqnarray}
		\bar{\mathbf{E}} \otimes \{\widehat{\mathbf{H}}(0)|\bar{\mathbf{E}}\} = \left\{\boldsymbol{e}_1 \otimes \{\widehat{\mathbf{H}}(0)|\bar{\mathbf{E}}\}\:\ \boldsymbol{|}\dots  \boldsymbol{|} \:\ \boldsymbol{e}_R \otimes \{\widehat{\mathbf{H}}(0)|\bar{\mathbf{E}}\}  \right\}, \notag
	\end{eqnarray}
	where,
	\setlength{\arraycolsep}{0pt}
	\begin{eqnarray}
		\boldsymbol{e}_i \otimes \{\widehat{\mathbf{H}}(0)|\bar{\mathbf{E}}\}&=&\left[\bigoplus_{r=1}^{R+1}(e\otimes h_{jkr}) \right]_{1\le j \le M,\; 1 \le k \le N}, \notag\\
		&=&\left[\bigoplus_{r=1}^{R+1}h_{jkr} \right]_{1\le j \le M,\; 1 \le k \le N}, \notag\\
		&=&\underset{r\in{1,\dots,R}}{\max}\{\mathbf{H}_r(0)|\bar{\mathbf{E}}\}, \notag \\
		&=&\mathbf{Q}^+(0). \notag \label{P_0} 
	\end{eqnarray}
	\setlength{\arraycolsep}{5pt}\noindent
	Similarly, we can obtain,
	\setlength{\arraycolsep}{0pt}
	\begin{eqnarray}
		&&-\left[\boldsymbol{e}_i \otimes -\{\widehat{\mathbf{H}}(0)|\bar{\mathbf{E}}\}\right]=\mathbf{Q}^-(0). \notag \label{N_0}
	\end{eqnarray}
	Therefore, Equation \eqref{H_t proof} is represented as:
	\setlength{\arraycolsep}{0pt}
	\begin{eqnarray}
		\widehat{\mathbf{H}}(t_*) &=& \left\{\mathbf{Q}^+(0)\ |\dots|\ \mathbf{Q}^+(0)\right\}
		+ \left\{\mathbf{Q}^-(0)\ |\dots|\ \mathbf{Q}^-(0)\right\}, \notag \\
		&=& \left\{\mathbf{Q}^+(0)+\mathbf{Q}^-(0)\ |\dots|\ \mathbf{Q}^+(0)+\mathbf{Q}^-(0)\right\}, \notag \\
		&=&\left\{\mathbf{H}_1(t_*)\ |\dots|\ \mathbf{H}_R(t_*)\right\}.
	\end{eqnarray}        
	This implies that:
	\setlength{\arraycolsep}{0pt}
	\begin{eqnarray}
		\mathrm{H}_i(t_*) &=& \mathrm{H}_j(t_*),\\
		&=&\mathbf{Q}^+(0) + \mathbf{Q}^-(0),\quad \forall i,j \in \mathcal{N}.
	\end{eqnarray}
	The result corresponds to \eqref{A7}$\sim$\eqref{A8} in \textbf{Definition~\ref{def4}}, and therefore, it indicates that the dual-extrema consensus is achieved.
\end{proof}

\bibliography{references.bib}

\end{document}